\documentclass[twocolumn,aps,prl]{revtex4-1}

\usepackage{extarrows}
\usepackage{amsmath}
\usepackage{graphicx,epstopdf}
\usepackage{bm}

\newcommand{\Qij}{\mathcal{Q}_{ij}}
\newcommand{\TT}{\mathbf{T}}
\newcommand{\II}{\mathbf{I}}
\newcommand{\TI}{\mathbf{TI}}

\usepackage[colorlinks=true, letterpaper=true, pdfstartview=FitV,
  linkcolor=blue, citecolor=blue, urlcolor=blue]{hyperref}

\begin{document}

\title{Orbital Magnetic Quadrupole Moment and Nonlinear Anomalous Thermoelectric Transport}

\author{Yang Gao}

\affiliation{Department of Physics, Carnegie Mellon University,
  Pittsburgh, PA 15213, USA}

\author{Di Xiao}

\affiliation{Department of Physics, Carnegie Mellon University,
  Pittsburgh, PA 15213, USA}

\date{\today}

\begin{abstract}
We present a microscopic theory of the magnetic quadrupole moment density $\Qij$ in periodic crystals with combined time reversal ($\TT$) and inversion ($\II$) symmetry. We obtain a gauge-invariant expression with clear physical interpretation and demonstrate the typical behaviour of $\Qij$ in a minimal two-band model that hosts a tilted Dirac cone.  We then show that $\Qij$ leads to an intrinsic nonlinear anomalous thermoelectric current.  As an example, we calculate the nonlinear Nernst and Hall current in the loop-current model for cuprate superconductors, and demonstrate their unique behaviour and capability of indicating $\TI$-invariance.
\end{abstract}

\maketitle

The classification of electronic states with broken symmetries is a fundamental issue in physics.  An interesting scenario arises when a state breaks time-reversal ($\TT$) and inversion ($\II$), but retains the combined $\TI$ symmetry.  Such a state can be realized by a vortex-like arrangement of spin magnetic moments in certain magnetoelectrics~\cite{Aken2007,Spaldin2008,Fiebig2016}, or antiferromagnetically aligned microscopic current loops as predicted in excitonic insulators~\cite{Halperin1968,Gorbatsevich1994} and the pseudogap regime of cuprate superconductors~\cite{Varma1997,Varma2006}.  These states are often called the hidden order phase because the $\TI$-invariance forbids a net macroscopic magnetization, making their experimental detection a challenging task~\cite{Simon2003,Shekhter2009,Orenstein2011,Matteo2012,Yakovenko2015}.

On the theory side, symmetry considerations have led to the proposal of the toroidization $\bm{\mathcal T}$ as the order parameter of $\TI$-invariant states~\cite{Spaldin2008,Dubovik1990}.
It is a time-odd polar vector derived from the antisymmetric part of the magnetic quadrupole moment density $\Qij$, i.e., $\mathcal T_k = \frac{1}{2}\epsilon_{ijk}\Qij$.  Classically, $\Qij$ is defined by~\cite{Dubovik1990}
\begin{equation}\label{eq_tclass}
\Qij =\frac{1}{3V} \int r_i (\bm r\times \bm J)_j \, d\bm r \;,
\end{equation}
where $\bm J$ is the current density.  However, a microscopic understanding of $\Qij$ in crystals has remained elusive.  This is because the position operator in Eq.~\eqref{eq_tclass} is ill-defined in the Bloch representation, the very same difficulty that has motivated the modern theory of electric polarization~\cite{King-smith1993,Resta1994} and orbital magnetization~\cite{Xiao2005,Thonhauser2005}.  Without a proper theory of $\Qij$, it is difficult to connect experimental observations to microscopic theories.

In this Letter, we bridge this gap by providing a microscopic derivation of the orbital $\Qij$ in periodic crystals.  The quantum mechanical theory of the spin part of $\Qij$ has been recently studied in Ref.~\cite{Gao2017}.
Using the semiclassical theory of electron dynamics~\cite{Xiao2010,Gao2014,Gao2015}, we obtain a gauge-invariant expression in terms of bulk Bloch functions.  The resulting $\Qij$ enjoys several desired properties: it satisfies all symmetry requirements, and its derivative with respect to chemical potential recovers the magnetoelectric polarizability in an insulator~\cite{Gao2017}.  Typical behavior of $\Qij$ is demonstrated in a minimal two-band model with a tilted Dirac cone.

Moreover, based on our microscopic theory, we show that $\Qij$ directly leads to a nonlinear \emph{intrinsic} anomalous thermoelectric current.  Since the linear anomalous current is forbidden in $\TI$-invariant systems, the nonlinear current is the leading order contribution and can be used to probe such symmetry.  As a concrete example, we calculate this nonlinear anomalous current in the loop-current model for cuprate superconductors~\cite{Varma1997,Varma2006}.  We find that both the nonlinear Hall and nonlinear Nernst effects are nonzero in this system, and they are greatly enhanced at the saddle points and the Dirac points of the energy bands, respectively.  Our result thus allows a more quantitative analysis of $\TI$-invariant systems beyond pure symmetry considerations.

{\it Microscopic derivation.}---To overcome the difficulty of the ill-defined position operator, we start by defining the quadrupole moment density $\Qij$ as a response function. We consider a homogeneous periodic crystal, perturbed by an inhomogeneous magnetic field $\bm B(\bm r)$. If the magnetic field as well as its spatial variation is small, $\Qij$ can be obtained from the local free energy density $F(\bm r)$ as follows:
\begin{equation}\label{eq_tdef}
\Qij(\bm r)=-\lim_{\bm B(\bm r)\rightarrow 0}\left. \frac{\partial F(\bm r)}{\partial (\partial_iB_j)}\right|_{\bm B(\bm r)}\,.
\end{equation}
Here in taking the derivative with respect to $\partial_i B_j$, the magnetic field at $\bm r$ needs to be kept fixed.

To carry out a perturbative calculation, we assume that the system can be described by a non-interacting, mean field Hamiltonian $\hat H$ that is commensurate with the crystal lattice.  The perturbed Hamiltonian can be written as $\hat{ H}_{\rm F}=\hat{ H}(\hat{\bm p}+\bm A(\bm r);\bm r)\,,$ where $\hat{\bm p}=i\bm \partial_{\bm r}$ is the momentum operator, and $\bm A(\bm r)$ is the magnetic vector potential.  For simplicity we have set $e = \hbar = 1$.  The vector potential is chosen to be
\begin{equation}\bm A(\bm r) = \frac{1}{2}\bm B(\bm r)\times \bm r
 - \frac{1}{6}(r_i\partial_i\bm B) \times \bm r \,.
\end{equation}
One can verify that $\bm A(\bm r)$ generates the correct magnetic field up to the first order derivative of $\bm B(\bm r)$.

With the above setup, we are ready to evaluate the correction to the free energy perturbatively.  Our tool of choice is the semiclassical theory of Bloch electron dynamics~\cite{Xiao2010,Gao2014, Gao2015}.  This approach has been used to calculate $\Qij$ due to the spin moments in Ref.~\cite{Gao2017}.  However, there is a notable difference.  For the spin $\Qij$, the magnetic field enters only through the Zeeman coupling, hence only first-order calculation is needed.  Here, the magnetic field enters through the minimal coupling, and we need to calculate the free energy response to the second order derivatives of $\bm A(\bm r)$.  Since the semiclassical formalism is by now a fairly standard approach, we leave the details of the derivation in the supplementary~\cite{suppl}, and only present the final result. For a single Bloch band labeled by $0$, $\Qij$ is given by
\begin{widetext}
\begin{equation}\label{eq_result}
\Qij = \int \frac{d\bm k}{(2\pi)^3} \Bigl\{
f\Bigl[\sum_{n\neq 0}\frac{2}{3}{\rm Re}[ (\mathcal{A}_i)_{0n} (M_j)_{n0}]-\frac{1}{12} \epsilon_{k\ell j}\partial_k(\Gamma_{i\ell})_0\Bigr]
+\mathcal{G}\Bigl[-\sum_{n\neq 0}\frac{2{\rm Re}[(\mathcal{ A}_i)_{0n} (M_j)_{n0}]}{\varepsilon_0-\varepsilon_n}-\frac{1}{3}\epsilon_{k\ell j}\partial_k g_{i\ell}\Bigr]\Bigr\} \;,
\end{equation}
\end{widetext}
where $n$ is the band index, $\varepsilon_0$ and $\varepsilon_n$ are band energies with $|u_0\rangle$ and $|u_n\rangle$ being the periodic part of the corresponding Bloch functions. $f$ is the Fermi-Dirac distribution function, and $\mathcal{G}=-k_BT\ln(1+\exp((\mu-\varepsilon_0)/k_BT))$ is the grand potential density. The zero temperature formula of $\Qij$ can be obtained by taking the $T \to 0$ limit, where $f = \Theta(\mu-\varepsilon)$ and $\mathcal{G} = (\varepsilon-\mu)\Theta(\mu-\varepsilon)$.
The derivatives in Eq.~\eqref{eq_result} are all with respect to the momentum $\bm k$. Einstein summation convention is implied for repeated indices.

Even though Eq.~\eqref{eq_result} has a rather complicated appearance, each term has a clear physical meaning.  The quantity $\bm{\mathcal A}_{0n}$, defined by $\bm{\mathcal A}_{0n}=\langle u_0|i\bm \partial_{\bm k}|u_n\rangle$, is the Berry connection.  It also has the meaning as the interband element of the position operator~\cite{Xiao2010}.  The quantity $\bm M_{n0}$, defined by $\bm M_{n0} = \sum_{m\neq 0}\frac{1}{2}(\bm v_{nm}+\bm v_0\delta_{nm})\times \bm {\mathcal A}_{m0}$ with $\bm v_{nm}=\langle u_n|\hat{\bm v}|u_m\rangle$ the velocity matrix element, thus has the meaning of the interband orbital magnetic moment.  Together, the first term in Eq.~\eqref{eq_result} can be interpreted as the quantum mechanical counterpart of the classical definition $\bm r_i(\bm r\times \bm J)_j$ in Eq.~\eqref{eq_tclass}.  In the second term, $(\Gamma_{js})_0=\langle u_0|\partial_j \hat{v}_s|u_0\rangle$ is the Hessian matrix. We note that $\partial_j \hat{v}_s$ represents the noncommutativity between $\bm r$ and $\bm v$. Therefore, the second term in Eq.~\eqref{eq_result} simply accounts for the fact that $r_i$ and $(\bm r\times \bm J)_j$ is noncommutative.

Next we turn to the third and fourth term in Eq.~\eqref{eq_result}. The third term is due to the positional shift of the electron wave packet under a magnetic field~\cite{Gao2014}. The same term also appears in the spin toroidization derived in Ref.~\cite{Gao2017}, with the orbital magnetic moment $\bm M_{n0}$ replaced by the spin magnetic moment.  In the last term, $g_{ij}=\sum_{n\neq 0}{\rm Re}[(\mathcal{A}_i)_{0n}(\mathcal{A}_j)_{n0}]$ is the quantum metric for band $0$. The appearance of the quantum metric is characteristic of the second order semiclassical theory~\cite{Gao2014,Gao2015}. Different from the conventional perturbative term through interband virtual transition as in the previous term, it represents the mixing of Bloch states within the same Bloch band~\cite{Gao2015}.

The expression of $\Qij$ in Eq.~\eqref{eq_result} satisfies several general requirements.  First, it is gauge-invariant, as an arbitrary phase factor added to $|u_0\rangle$ does not yield any change to Eq.~\eqref{eq_result}.  It can be also easily generalized to the multi-band case by summing over all the occupied bands, instead of the single band $0$.

Secondly, as dictated by general thermodynamic principles~\cite{Gao2017}, $\Qij$ should be related to the orbital magnetoelectric polarizability $\alpha_{ij} \equiv dP_i/dB_j$ in insulators, where $P_i$ is the polarization induced by the magnetic field $B_j$.  Set $T = 0$ and differentiate $\Qij$ with respect to $\mu$.  The first two terms multiplied by $f$ vanishes for an insulator. The forth term yields a total derivative, which again vanishes for an insulator. By comparing the third term with the expression of $\alpha_{ij}$ obtained in Ref.~\cite{Essin2010}, we find
\begin{equation}\label{eq_connt}
e\frac{\partial \Qij}{\partial \mu}=-\alpha_{ij} \;.
\end{equation}
The same relation has also been obtained by Shitade \textit{et al.}~\cite{Shitade2018}.

Finally, we check the transformation properties of $\Qij$.  Due to the appearance of the three $\bm k$-derivatives in each term, Eq.~\eqref{eq_result} is manifestly odd under $\TT$ and $\II$, but even under $\TI$. As for point group operations, note that the magnetic moment $\bm M$ transforms as an axial vector and the Berry connection $\bm{\mathcal A}$ transform as a polar vector, indicating that the two terms containing $\bm M$ and $\bm{\mathcal A}$ are pseudotensors. The remaining terms share the same property due to the appearance of the Levi-Civita symbol. Therefore, $\Qij$ should transform as a pseudotensor, whose antisymmetric part corresponds to a polar vector, i.e. the orbital toroidization $\bm {\mathcal T}$, under point group operations.

{\it Minimal model.}---To demonstrate our theory, we consider a minimal two-band model with the following Hamiltonian,
\begin{equation}\label{eq_hamil1}
\hat{H}= v^\prime k_x + v_xk_x\sigma_x+v_y k_y\sigma_y+\Delta \sigma_z\,,
\end{equation}
where $\sigma_x$, $\sigma_y$ and $\sigma_z$ are Pauli matrices, $v^\prime$, $v_x$, $v_y$ are characteristic velocities, and $\Delta$ is the energy gap. The energy spectrum has the shape of a tilted Dirac cone, with $\varepsilon_{\pm}=v^\prime k_x\pm (v_x^2k_x^2+ v_y^2k_y^2+\Delta^2)^{1/2}$.  In practice, $v^\prime$ can be tuned by applying strain to systems that host Dirac points.

The tilting is essential to have a finite $\Qij$.  To see this, we note that for two dimensional systems the electrons can only couple to $B_z$. Therefore, the only non-trivial components of $\Qij$ are $\mathcal{Q}_{xz}$ and $\mathcal{Q}_{yz}$.  Since $\mathcal{Q}_{xz} = -\partial F/\partial (\partial_x B_z)$ and the system is invariant under the mirror-$y$ operation, $\mathcal{Q}_{xz}$ must vanish.  On the other hand, the tilting $v'k_x$ breaks the mirror-$x$ symmetry, and hence a nonzero $\mathcal{Q}_{yz}$ is allowed. We note that the analysis based on mirror symmetries also applies to three-dimensional tilted Dirac cones, and a finite $\Qij$ is thus expected for Dirac/Weyl semimetals that breaks both $\TT$ and $\II$, and that has a finite tilting.

\begin{figure}[t]
  \includegraphics[width=\columnwidth]{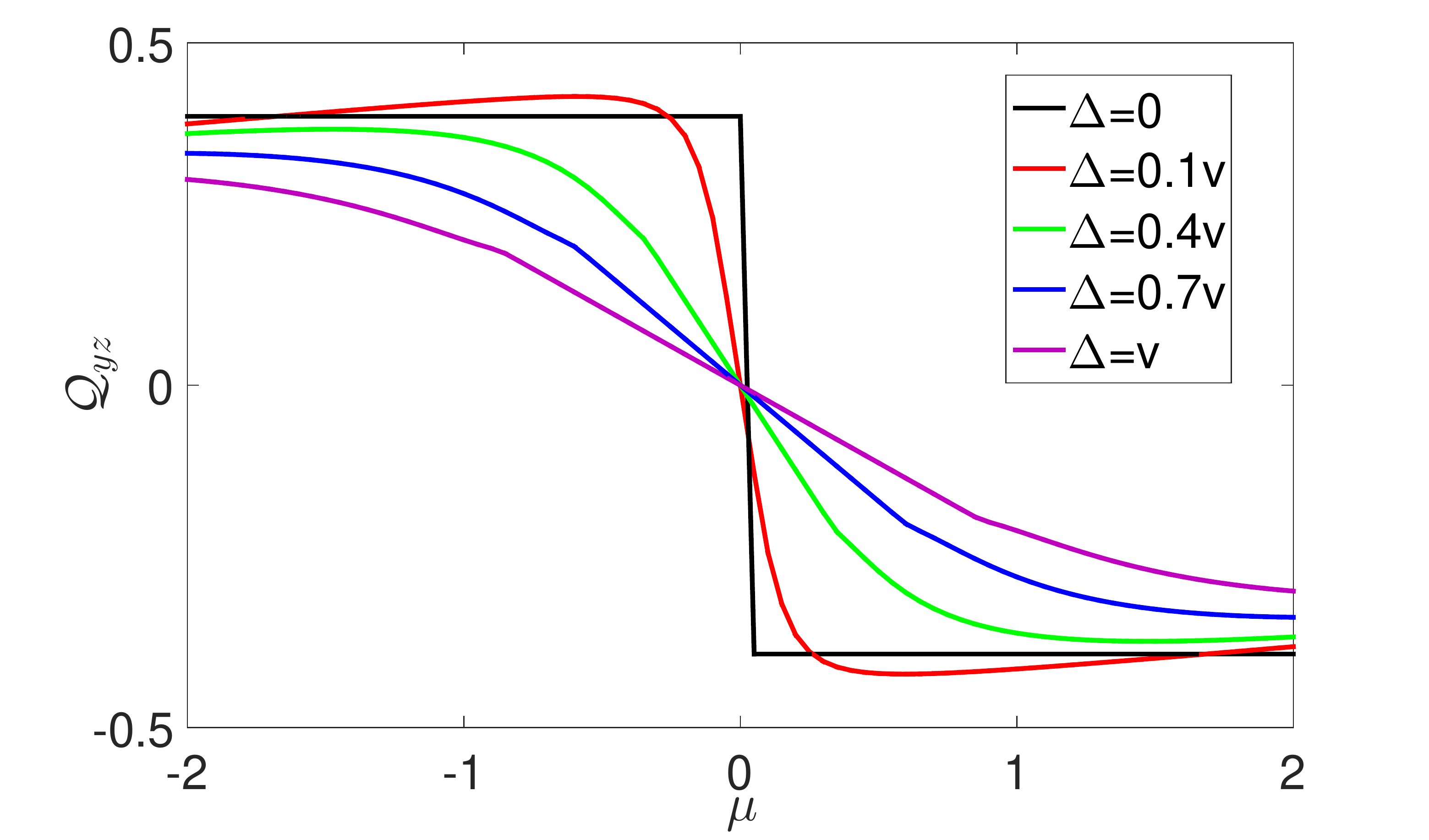}
  \caption{$\mathcal{Q}_{yz}$ for two-dimensional tilted Dirac cone. The parameter is chosen as follows: $v_x=v_y=v$, $v^\prime=0.5v$. The chemical potential is in unit of $v$. $\mathcal{Q}_{yz}$ is in unit of $ev/4\pi^2$.}\label{fig_model1}
\end{figure}

To calculate $\mathcal{Q}_{yz}$ we first set $\Delta =0$, $T=0$, and assume that the chemical potential falls in the valance band ($\mu<0$). We find that $\mathcal{Q}_{yz}$ is a constant $\frac{ev^\prime} {16\pi}\frac{|v_y|}{|v_x|}$~\cite{suppl}. If the Fermi energy falls in the conduction band, similar calculation shows that $\mathcal{Q}_{yz}$ is the same constant with an opposite sign: $-\frac{ev^\prime} {16\pi}\frac{|v_y|}{|v_x|}$. Therefore, $\mathcal{Q}_{yz}$ experiences a finite jump when the chemical potential crosses the Dirac point.  The magnitude of the jump is proportional to the tilting near the Dirac point, consistent with the above symmetry analysis.  This sudden jump behavior is a peculiar property of the Dirac point physics. As we can see below, as soon as a gap opens up at the Dirac points, the curve becomes smooth.  When the gap is finite, we resort to direct numerical integration of Eq.~\eqref{eq_result}.  Figure~\ref{fig_model1} shows $\mathcal{Q}_{yz}$ as a function of the gap parameter $\Delta$.  It is obvious that as the gap becomes smaller, $\mathcal{Q}_{yz}$ behaves closer to the Heaviside function.  As the chemical potential moves deeper in the band, the orbital toroidization approaches the same value independent of $\Delta$.

{\it Intrinsic nonlinear transport.}---With the microscopic theory of $\Qij$ fully established, we now discuss its consequences in transport experiments.  According to the theory of macroscopic electromagnetism~\cite{Jackson,deGroot,Dubovik1990}, an inhomogeneous medium can carry an equilibrium current due to magnetization $\bm M$ and magnetic quadrupole moment $\Qij$:
$\bm J_\text{medium}=\bm \nabla\times \bm M-\bm \nabla\times (\partial_i\mathcal{Q}_{ij} \hat{\bm e}_j)$.  In transport studies, this current must be discounted since it cannot be measured by conventional transport experiments.  This point has been extensively discussed in the literature~\cite{Cooper1997,Xiao2006}.  Consequently, the current measured in a transport experiment is given by
\begin{align}\label{eq_current}
\bm J=\bm J_\text{local}-\bm \nabla\times \bm M+\bm \nabla\times (\hat{e}_j\partial_i \Qij)\,,
\end{align}
where $\bm J_\text{local}=-e{\rm Tr}[\hat{\bm v}\delta(\bm r-\hat{\bm r})]$ is the local current density.  Equation~\eqref{eq_current} is valid when local equilibrium can be established everywhere in the system.

The first order current due to $\bm\nabla\times\bm M$ has been calculated in Ref.~\cite{Xiao2006}.  Here with our microscopic theory of $\Qij$ and by further calculating $\bm J_\text{local}$ up to second order, we can obtain the second order current.  For illustration purposes, we assume that the spatial inhomogeneity is induced by a constant temperature gradient $\bm \nabla T$.  Leaving the details in the supplementary~\cite{suppl},  we obtain the corresponding thermoelectric current at second order,
\begin{equation}\label{eq_thermo}
\bm J^{(2)} = -\int \frac{d\bm k}{(2\pi)^3}\bm \nabla T\times(\hat{e}_i\theta_{ij}\partial_jT) \frac{(\varepsilon_0-\mu)^2}{T^2}\frac{\partial f}{\partial \mu}\,,
\end{equation}
where $\theta_{ij}=2{\rm Re}\sum_{n\neq 0}[(\mathcal A_i)_{0n}(\bm v_0\times \bm {\mathcal A}_{n0})_j/(\varepsilon_0-\varepsilon_n)]$.

The current $\bm J^{(2)}$ is independent of the relaxation time, and is therefore of intrinsic nature. It also flows perpendicular to the temperature gradient in the absence of a magnetic field, i.e., it is an anomalous current.  Moreover, if the system respects $\TI$ symmetry, the linear order anomalous current in Eq.~\eqref{eq_current} always vanishes, leaving the nonlinear current $\bm J^{(2)}$ as the leading order contribution.  The existence of an intrinsic nonlinear Hall-type current combined with the lack of its linear counterpart can thus serve as a signature of $\TI$-invariance.

We have also derived the current driven by the chemical potential gradient at second order~\cite{suppl}. By comparing this current with Eq.~\eqref{eq_thermo}, we find that it shares a structural similarity with Eq.~\eqref{eq_thermo} and can be obtained from Eq.~\eqref{eq_thermo} by replacing $(\varepsilon_0-\mu)(\bm \nabla T/T)$ with $\bm \nabla\mu$. Since $\bm \nabla \mu$ is equivalent to an electric field $\bm E$, according to the Einstein relation, our current should coincide with the nonlinear anomalous Hall current driven by $\bm E$. This latter current has been derived in a completely different setting~\cite{Gao2014}. By comparing these two currents, we confirm their coincidence and hence the validity of our result.

Recently, an extrinsic nonlinear Hall effect has been discussed in Ref.~\cite{Inti2015}, in which the nonlinear current is linearly proportional to the relaxation time.  However, it vanishes in $\TI$-invariant systems.  In systems with broken $\TT$, $\II$, and $\TI$, the extrinsic and intrinsic contribution coexist, but the intrinsic one would dominate in dirty samples where the relaxation time is small.

\begin{figure}[t]
  \includegraphics[width=\columnwidth]{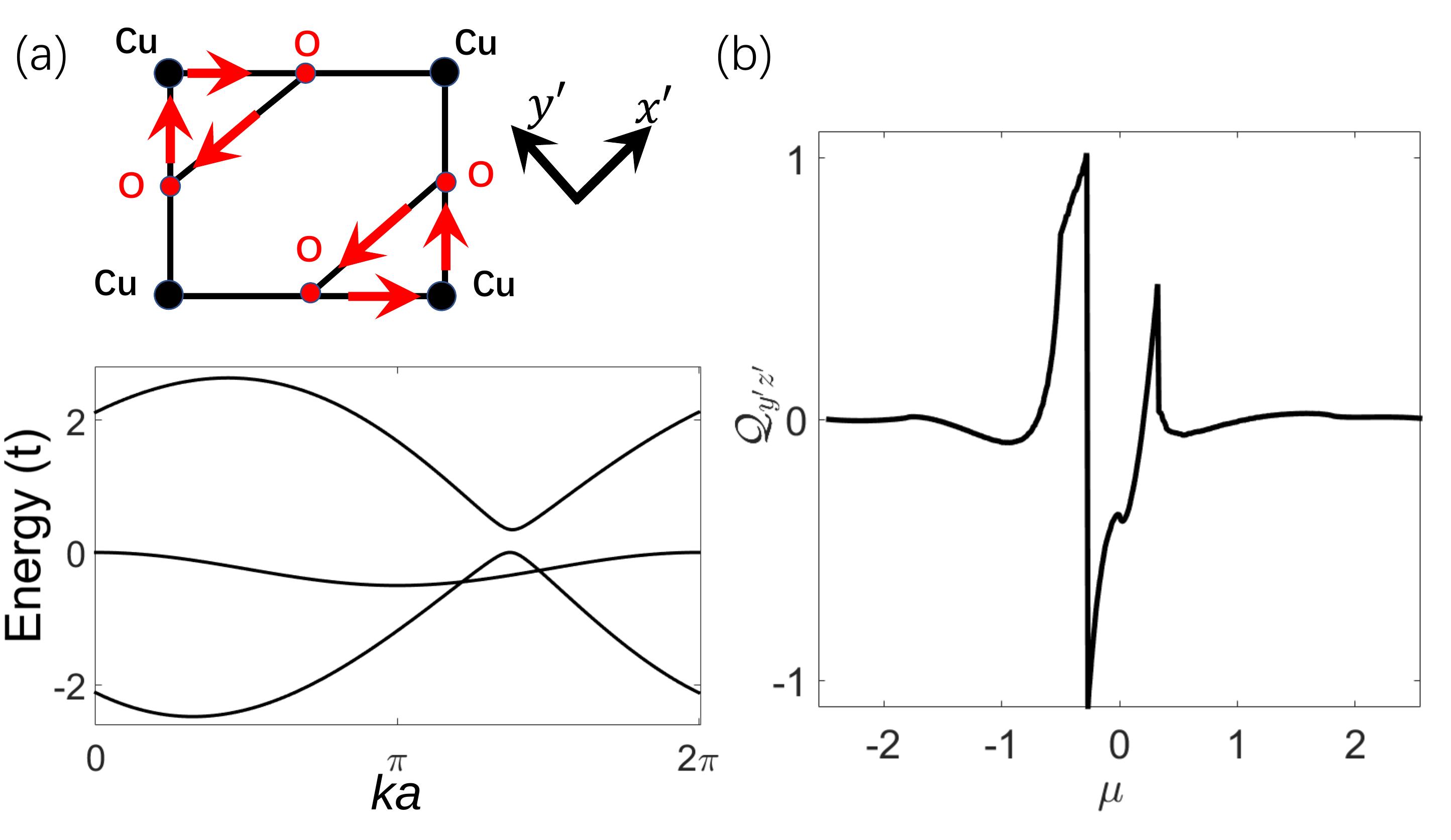}
  \caption{Loop-current state and spectrum (left panel) and $\mathcal{Q}_{y^\prime z^\prime}$ (right panel). In panel (a), the red arrows show the direction of the microscopic current. The energy spectrum is plotted along $(1,1,0)$ direction. The parameter choices are $r=1.5t$ and $t^\prime=0.5 t$. $\mu$ and $\mathcal{Q}_{y^\prime z^\prime}$ are in units of $t$ and $eat/16\pi^2\hbar$, respectively.}
\label{fig_fig2}
\end{figure}

{\it Loop-current state.}---As a concrete example, we consider the loop-current model proposed for the pseudogap regime of cuprate superconductors~\cite{Varma1997,Varma2006}.  In this model, microscopic current loops develop within the Cu-O plane due to strong Coulomb interactions.  Depending on their arrangement, the system can be $\TI$-invariant.  Figure~\ref{fig_fig2}a shows such a state: the intracell current loops break $\TT$, but each cell consists of two antiferromagnetically aligned current loops, which retains the combined $\TI$ symmetry. As a result, the linear anomalous current vanishes, but the nonlinear anomalous current should exist.  In the following we calculate the magnetic quadrupole density and nonlinear thermoelectric transport coefficients of this model using our theory.

The Hamiltonian of the loop-current model shown in Fig.~\ref{fig_fig2}a is given by~\cite{He2012}
\begin{equation}\label{eq_model}
\hat{H}=\begin{pmatrix}
0 & its_x+irc_x & its_y+irc_y\\
-its_x-irc_x &0&t^\prime s_xs_y\\
-its_y-irc_y & t^\prime s_xs_y &0
\end{pmatrix}\,,
\end{equation}
where $s_x=\sin(k_xa/2)$, $c_x=\cos(k_xa/2)$, and $a$ is the lattice constant. The basis of the above model is $(|d\rangle,|p_x\rangle,|p_y\rangle)$ where the $d$-orbital is from copper and the two $p$-orbitals are from oxygen. $t$ and $t^\prime$ are the hopping strengths for copper-oxygen and oxygen-oxygen from the kinetic energy. $r$ is the effective hopping strength between copper and oxygen from the Coulomb interaction.

This model has three bands. With our choice of parameters the lower two bands touch linearly at two points as shown in Fig.~\ref{fig_fig2}a, with energies at $-0.44t$ and $-0.27t$ respectively. The upper two bands also touch at two points, but at the same energy $0.33t$ (these points are not on the high symmetry lines and are thus not shown in Fig.~\ref{fig_fig2}a).

Before calculating $\Qij$, we first analyze the symmetry of this model.  Let us define the $x'$ axis along the $[1,1,0]$ direction and the $y'$ axis along the $[1,\bar{1},0]$ direction as indicated in Fig.~\ref{fig_fig2}a.  The model Hamiltonian in Eq.~\eqref{eq_model} breaks the mirror-$x'$ symmetry but preserves the mirror-$y'$ symmetry.  Following our discussion earlier, the only nonzero component of $\Qij$ is $\mathcal{Q}_{y'z'}$. 

\begin{figure}[t]
  \includegraphics[width=\columnwidth]{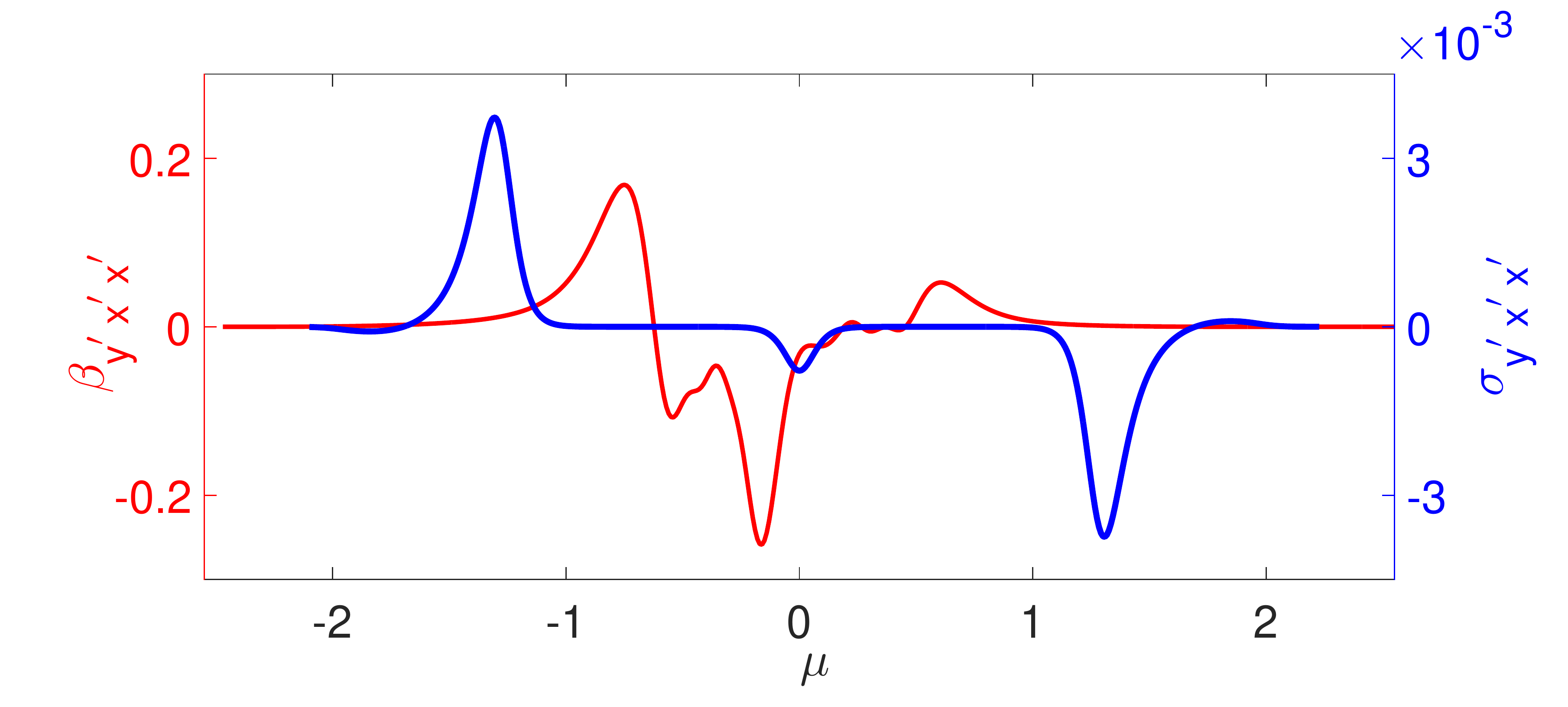}
  \caption{Nonlinear thermoelectric (red) and nonlinear electric (blue) conductivity for the model in Eq.~\eqref{eq_model}. $\beta_{x^\prime y^\prime y^\prime}$ and $\sigma_{x^\prime y^\prime y^\prime}$ are in units of $et/(16\pi^2 a\hbar)$ and $e/(16\pi^2 ta\hbar)$, respectively. Both quantities are calculated using self-adaptive method. We also set a finite temperature at $T=0.05t$, and add a small imaginary part $0.001i$ to the denominator of $\bm \theta$ in Eq.~\eqref{eq_thermo}. }\label{fig_fig3}
\end{figure}

Figure~\ref{fig_fig2}b shows the calculated $\mathcal{Q}_{y^\prime z^\prime}$ as a function of chemical potential. The peak structure can be traced back to the finite jump near the three tilted Dirac points located at $\mu = -0.44t$, $-0.27t$, $0.33t$.  A detailed analysis is presented in the Supplmenetary~\cite{suppl}.  In Fig.~\ref{fig_fig3},  we plot the nonlinear anomalous thermoelectric conductivity $\beta$ and the nonlinear anomalous electric conductivity $\sigma$ as a function of the chemical potential. They can be obtained from the nonlinear current through the relation $J_i = \beta_{ijk}(\partial_jT\partial_kT)/T^2 + \sigma_{ijk}(\partial_j \mu \partial_k \mu)$.  Due to the mirror-$y'$ symmetry, the only nonzero off-diagonal components are $\beta_{x^\prime y^\prime y^\prime}$ and $\sigma_{x^\prime y^\prime y^\prime}$.

Both response functions are very sensitive to the Fermi energy.  However, $\beta_{x^\prime y^\prime y^\prime}$ and $\sigma_{x^\prime y^\prime y^\prime}$ has different peak structures as shown in Fig.~\ref{fig_fig3}. The nonlinear thermoelectric conductivity varies drastically near the three Dirac points, in accordance with the peak structure in $\mathcal{Q}_{y^\prime z^\prime}$.  In comparison, the nonlinear electric conductivity has peaks in different energy ranges, far away from those Dirac points. The reason is as follows. Besides the three Dirac points, the model Hamiltonian also has three saddle points, at around $\mu=0$ and $\mu=\pm 1.8 t$.  Near the saddle points, the density of states diverges. The peak structure of $\sigma$ is exactly due to this divergent density of states. The expression of $\beta$ contains a factor $(\varepsilon-\mu)^2$, which cancels the divergent behavior of the density of states. As a result, $\beta$ does not show peaks near the saddle points.

In addition to the nonlinear transport experiment proposed here, other techniques such as neutron scattering~\cite{Mangin2015,Fauque2006,Jeong2017}, X-ray diffraction~\cite{Scagnoli2011}, and second harmonic generation~\cite{Zhao2015,Zhao2016} have been applied to probe possible loop-current states in various materials, such as YBCO, CuO, and Sr$_2$IrO$_4$.  So far the analysis has been mostly limited to symmetry considerations~\cite{Matteo2012,Simon2003,Shekhter2009}.  A future direction is to apply our theory of $\Qij$ to these systems, which may shed new light on the microscopic origin of the loop-current states.  We should also mention that even though we have mainly considered the loop-current model as an example, our theory is applicable to any $\TI$-invariant systems.




\begin{acknowledgments}
This work is supported by the Department of Energy, Basic Energy
Sciences, Grant No.~DE-SC0012509.
\end{acknowledgments}

\end{document}